\documentclass[twocolumn,aps,prl]{revtex4-1}
\usepackage[usenames,dvipsnames]{xcolor}
\usepackage{times}
\usepackage{amsmath}
\usepackage{amsfonts}
\usepackage{amssymb}
\usepackage[colorlinks=true,citecolor=blue,linkcolor=red]{hyperref}
\usepackage{soul}
\usepackage{graphicx}
\usepackage{bbold}					
\usepackage[makeroom]{cancel}		
\usepackage{multirow}				
\usepackage[normalem]{ulem}        
\usepackage{array}
\usepackage{pdfpages}
\makeatletter
\AtBeginDocument{\let\LS@rot\@undefined}
\makeatother
\newcolumntype{x}[1]{>{\centering\let\newline\\\arraybackslash\hspace{0pt}}p{#1}}

\DeclareMathAlphabet{\mathbbold}{U}{bbold}{m}{n}

\newcounter{subeqn} %
\renewcommand{\Re}{\operatorname{Re}}
\renewcommand{\Im}{\operatorname{Im}} 
\makeatletter
\@addtoreset{subeqn}{equation}
\makeatother

\renewcommand{\Re}{\operatorname{Re}}
\renewcommand{\Im}{\operatorname{Im}}
\newcommand{\clr}{\color{red}}

\begin{document}
	
	\title{Two-dimensional Asymptotic Generalized Brillouin Zone Theory}
	
	\author{Zeqi Xu$^1$}\thanks{These two authors contributed equally}
	\author{Bo Pang$^1$}\thanks{These two authors contributed equally}
	\author{Kai Zhang$^2$}	
	\email[Corresponding author: ]{phykai@umich.edu}
	\author{Zhesen Yang$^1$}
	\email[Corresponding author: ]{yangzs@xmu.edu.cn}
	
	\affiliation{$^1$Department of Physics, Xiamen University, Xiamen 361005, Fujian Province, China}
	\affiliation{$^2$Department of Physics, University of Michigan, Ann Arbor, Michigan 48109, United States}
	
	\date{\today}
	
	\begin{abstract}
		In this work, we propose a theory on the two-dimensional non-Hermitian skin effect by resolving two representative minimal models. 
		Specifically, we show that for any given non-Hermitian Hamiltonian, (i) the corresponding region covered by its open boundary spectrum on the complex energy plane should be independent of the open boundary geometry; and (ii) for any given open boundary eigenvalue $E_0$, its corresponding two-dimensional asymptotic generalized Brillouin zone is determined by a series of geometry-independent Bloch/non-Bloch Fermi points and geometry-dependent non-Bloch equal frequency contours that connect them. 
		A corollary of our theory is that most symmetry-protected exceptional semimetals should be robust to variations in OBC geometry. 
		Our theory paves the way to the discussion on the higher dimensional non-Bloch band theory and the corresponding non-Hermitian bulk-boundary correspondence.  
	\end{abstract}
	
	\maketitle
	
	{\color{red} \em Introduction.}---The non-Hermitian skin effect (NHSE), a phenomenon characterized by the exponential localization of nearly all eigenstates at the boundary,
	has recently attracted considerable attention~\cite{Yao2018,Kunst2018PRL,Torres2018,ChingHua2019,LeeCH2019_PRL,Longhi2019_PRR,Slager2020PRL,Kai2020PRL,Okuma2020_PRL,YYFPRL2020,LiLH2020_NC,Kawabata2020,Yokomizo2020PRRB,Titus2020,Yoshida2020,Zirnstein2021PRL,CXGuo2021PRL,SunXQ2021PRL,BLZhangNC2021,LuMing2021,WangKai2021Science,Longhi2022PRL,SiboG2022PRA,XJLiu2023PRB,DingKun2022NRP,LeeCHReview,YFChen2022Review}. 
	From a physical perspective, the NHSE is highly counter-intuitive, because in systems with either discrete or continuous translational symmetry, it is generally expected that the open boundary condition (OBC) eigenstates should manifest as extended Bloch or plane waves. 
	When the Hamiltonian is Hermitian, this is indeed the case. 
	For instance, in Hermitian wave chaotic systems~\cite{ChaosRMP2015}, although the complexity of the geometric shape may influence the periodic patterns of the corresponding eigenstates' wave function, it cannot induce any localization behavior around the boundary.
	However, this conventional understanding fails when the Hamiltonian becomes non-Hermitian, leading to the emergence of the NHSE. 
	
	From a theoretical viewpoint, the development of a theory that accurately describes these exponentially localized skin modes is a crucial starting point for further research~\cite{SongFei2019,Ashvin2019PRL,Thomale2020,Ghatak2020,XuePeng2020,NeupertETI2020,Ueda2021PRL,XueWT2021PRB,Sato2021PRL,LiuYX2021_PRB,Fang2023PRB,XWT2022PRL,LQPRL2022,Nori2019PRL,LLHu2020PRL,SPKou2020PRB,XDZhang2021NC,CYF2021NC,XuePeng2021PRL,ChenWei2022PRB}. 
	For one-dimensional (1D) systems, this can be accomplished by the generalized Brillouin zone (GBZ) theory, which allows for the analytical calculation of both the OBC eigenvalues and eigenstates in the $N\rightarrow\infty$  limit~\cite{Yao2018,Murakami2019PRL,ZSaGBZPRL,Zhesen2020,GBloch2017PRB,DengTS2019,KawabataPRB2020,Kunst2021RMP,YMHu2021,OkumaReview2023}. 
	However, for two-dimensional (2D) systems, despite some attempts to establish a corresponding GBZ theory~\cite{WangZhong2018,Kai2022NC,Yokomizo2022PRB,HYWang2022,HPHu2023,HuiJiang2022}, a universal framework remains elusive. 
	
	The challenges in establishing the 2D GBZ theory may arise from several factors. 
	Firstly, unlike the 1D case, the characteristic equation (ChE) contains three complex variables $E_0$, $k_{x}$ and $k_y$. 
	Consequently, even for a given $E_0$, we are left with two complex variables and one complex constraint equation, whose solution is a 2D Riemann surface in the 4D space defined in $(\Re k_x,\Im k_x, \Re k_y, \Im k_y)$, making it challenging to handle both analytically and numerically.
	Secondly, the shape of OBC geometry can significantly affect the NHSE in two and higher dimensions~\cite{Kai2022NC,DDS2023PRL,Wang2022NC,KunDing2022NanoP,QYZhouN2023C,WanSciB,KunDarPRL2023,YQarXiv2023,Kai2023arXiv}. 
	For instance, in the geometry-dependent skin effect~\cite{Kai2022NC}, the NHSE disappears in square geometry but emerges in triangle geometry. 
	This suggests that the 2D GBZ theory, if it exists, will be highly dependent on the choice of the OBC geometry. 
	Therefore, even for a common Hamiltonian, one would need to study the 2D GBZ case by case, according to their corresponding OBC geometry shapes. 
	Given these considerations, generalizing the 2D GBZ theory poses a significant challenge for theoretical physics.
	
	In this Letter, we establish an asymptotic GBZ theory for the 2D NHSE, fully addressing these two challenges. 
	We propose a vector field representation for the ChE, exactly mapping its solution to a vector field on the Brillouin zone (BZ) without loss of information. 
	Based on this representation, the GBZ problem can be reduced to the task of identifying the boundary-allowed regions on the 2D BZ. 
	To pinpoint these regions, we propose a {\em dynamical-duality method}, which allows us to precisely construct the eigenstates of the Hamiltonian with arbitrary open-boundary geometry. 
	We apply our method to two representative models and obtain the corresponding GBZ exactly. 
	The insights gained form these models leads us to propose an intuitive generalization on the 2D GBZ applicable to all types of 2D NHSE. 
	Our theory essentially answers the crucial question of whether the exceptional points or the bulk boundary correspondence in the non-Hermitian systems are robust to the variation of OBC geometry shapes. 
	
	\begin{figure*}[t]
		\begin{center}
			\includegraphics[width=1\linewidth]{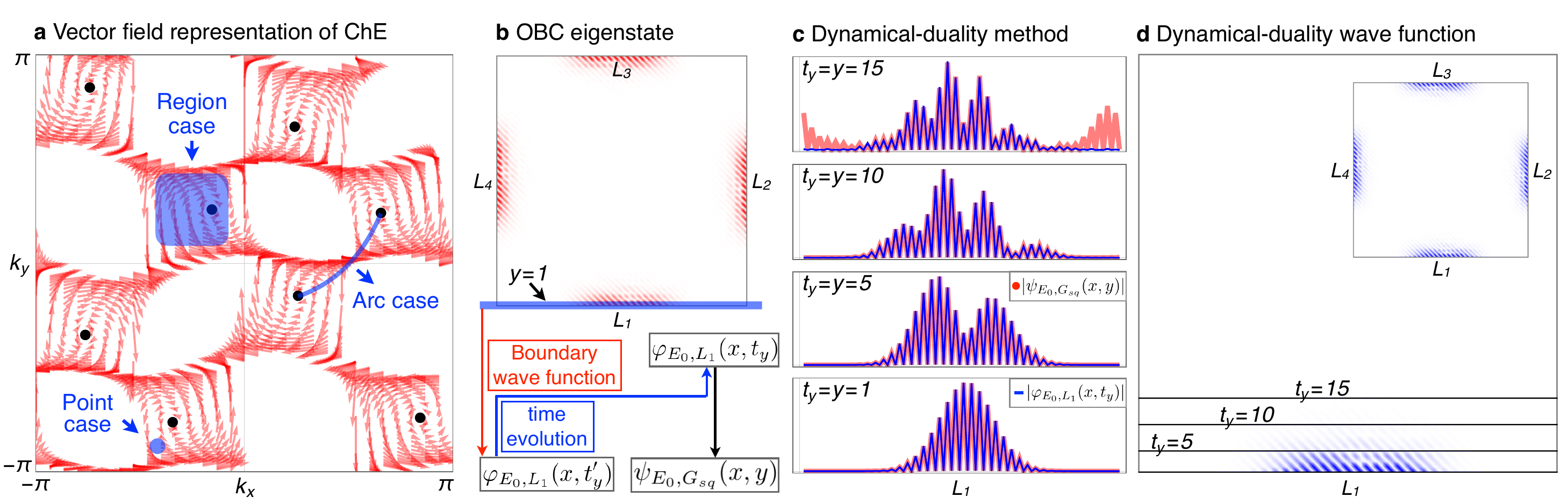}
			\par\end{center}
		\protect\caption{(a) shows the vector field representation of ChE. The Fermi points of $E_0$ are depicted as black points, while the red arrows signify the non-Bloch localization vectors $(k_x,k_y)\to(k_x+\mu_x,k_y+\mu_y)$. In this representation, the 2D GBZ can be categorized into point, arc, and region cases. (b) presents the OBC eigenstate $|\psi_{E_0,G_{sq}}(x,y)|$ under a square geometry with $N_x=N_y=80$. The wavefunction at the boundary $y=1$ is extracted and relabeled as $\varphi_{E_0,L_1}(x,t_y')$. (c) and (d) display the corresponding dynamics of $\varphi_{E_0,L_1}(x,t_y)$. In (c), the thick red and thin blue lines represent $|\psi_{E_0,G_{sq}}(x,y)|$ and $|\varphi_{E_0,L_1}(x,t_y)|$, respectively. In (d), the large and small squares represent $|\varphi_{E_0,L_1}(x,y)|$ and $|\varphi_{E_0,G_{sq}}(x,y)|$, respectively.} 
		\label{F1}
	\end{figure*}
	
	{\color{red} \em Vector-field representation of ChE.}---We begin with a general non-Bloch Hamiltonian  $H(z_x,z_y)$, where $(z_{x},z_y)\in\mathbb{C}^2$ represents the non-Bloch momentum and can be expressed as: 
	\begin{equation}
		z_x=e^{ik_x+\mu_x},~~~~z_y=e^{ik_y+\mu_y}.
		\label{z1}
	\end{equation}
	Here $k_{x/y}$ is the (real) Bloch momentum, and $\mu_{x/y}$ indicates the corresponding localization factor. 
	For a generic energy $E_0\in\mathbb{C}$, we can define the set of solutions of the ChE as: 
	\begin{equation}
		F_H(E_0):=\{(z_x,z_y)\in\mathbb{C}^2|\det[E_0-H(z_x,z_y)]=0\}.
		\label{FH}
	\end{equation} 
	Based on Eq.~\ref{z1}, the generalized momentum $(z_x,z_y)$ can be exactly mapped to a vector on the BZ, which starts at $(k_x,k_y)$ and ends at $(k_x+\mu_x,k_y+\mu_y)$, expressed as:
	\begin{equation}
		(z_x,z_y)\mapsto {\rm Vector:}~ (k_x,k_y)\rightarrow (k_x+\mu_x,k_y+\mu_y).  
	\end{equation}
	The length and direction of this vector represent the localization strength and direction, respectively, of the non-Bloch wave $|z_x,z_y\rangle$~\cite{SM1}.
	Due to the presence of the ChE, $\mu_{x}$ and $\mu_y$ are functions of $(k_{x},k_y$), which implies that when $(k_x,k_y)$ is given, $(\mu_x,\mu_y)$ will be determined. 
	Mapping all the solutions of the ChE to the vectors on the BZ, we can obtain a vector-field representation for $F_H(E_0)$. 
	Fig.~\ref{F1}(a) shows an example for the following Hamiltonian:
	\begin{equation}
		H_1(z_x,z_y)=-z_xz_y+z_x/z_y+z_y/z_x+1/(z_xz_y).
		\label{H1}
	\end{equation}
	with $E_0=0.52 - 0.64 i$ that belongs to both the PBC and OBC spectra. 
	The black dots in Fig.~\ref{F1}(a) represent the Fermi points of $E_0$ satisfying $(\mu_x,\mu_y)=(0,0)$. 
	In Appendix, we show that the GBZ of $E_0$ must be a subset of $F_{H_1}(E_0)$, and therefore, can be classified by the region, arc, and point cases as shown in Fig.~\ref{F1}(a).  
	Once the base field $(k_x,k_y)$ on the BZ are determined, the corresponding 2D GBZ of $E_0$ is almost determined. 
	
	\begin{figure*}[t]
		\begin{center}
			\includegraphics[width=1\linewidth]{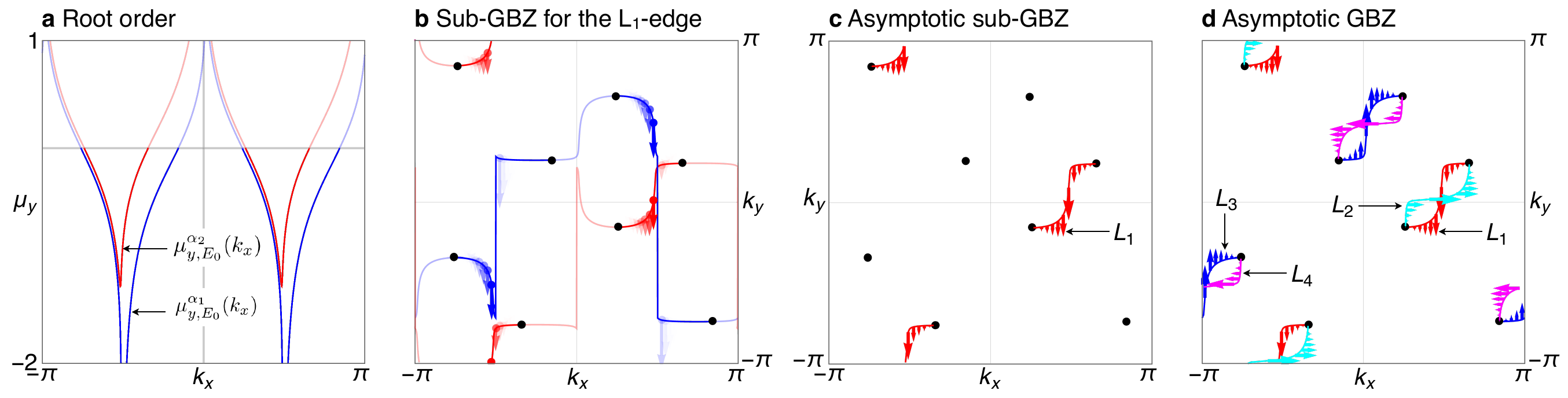}
			\par\end{center}
		\protect\caption{(a) depicts the root order. The blue and red lines symbolize  $\mu_{y,E_0}^{\alpha_1}(k_x)$ and $\mu_{y,E_0}^{\alpha_2}(k_x)$, respectively. Their parts that are less than zero are represented by opaque lines. The gray points denote $|\phi_{E_0,L_1}(k_x,t_y')|$ appeared in Eq.~\ref{ck}. (b) presents the vector field representation for the sub-GBZ of the $L_1$-edge. The black points indicate the Fermi points. The blue and red lines symbolize $k_{y,E_0}^{\alpha_1}(k_x)$ and $k_{y,E_0}^{\alpha_2}(k_x)$, respectively. The opaque curves represent the one satisfying Eq.~\ref{mu}. The color intensity of the points/arrows represents the values of $|A_{E_0,L_1}^{\alpha_{in}}(k_x)|$ that appear in Eq.~\ref{phinew}. (c) displays the asymptotic sub-GBZ of the $L_1$-edge. Here the black points denote the Fermi points. The arcs symbolize the non-Bloch equal frequency contours, i.e. $(k_x,k_y)=(k_x,k_{y,E_0}^{\alpha_2}(k_x))$, and the arrows indicate the non-Bloch localization vector $(\mu_x,\mu_y)=(0,\mu_{y,E_0}^{\alpha_2}(k_x))$. (d) presents the asymptotic GBZ of the square geometry. The asymptotic sub-GBZ for each edge is represented by different colors.} 
		\label{F2}
	\end{figure*}
	
	{\color{red} \em Dynamical-duality wavefunction.}---
	For the eigenvalue $E_0$ selected in Fig.~\ref{F1}(a), the corresponding OBC eigenstate under a square geometry $G_{sq}$, labeled by $\psi_{E_0,G_{sq}}(x,y)$, is plotted in Fig.~\ref{F1}(b) with the system size $N_x=N_y=80$. 
	Here, the transparency in red corresponds to the amplitude of the wavefunction $|\psi_{E_0,G_{sq}}(x,y)|$. 
	It is shown that the wavefunction is localized at four edges, labeled $L_{1,2,3,4}$, respectively.
	Now we construct $\psi_{E_0,G_{sq}}(x,y)$ analytically. 
	First, let's focus on the $L_1$-edge. 
	As shown in the flowchart at the bottom of Fig.~\ref{F1}(b), our central observation is that we can map our problem to a dynamical process. 
	Specifically, suppose that we have an initial wavefunction at the $L_1$-edge, which is labeled by $\varphi_{E_0,L_1}(x,t_{y}')$.
	Let this initial wavefunction evolve to some time $t_y$, resulting in $\varphi_{E_0,L_1}(x,t_y)$. 
	If we can judiciously choose a suitable dynamical evolution equation (which is detailed in Appendix), $\varphi_{E_0,L_1}(x,t_y)$ will reveal the information of the OBC eigenstate involving the $L_1$-edge. 
	
	As shown in Fig.~\ref{F1}(c), we compare the dynamical-duality wavefunction $|\varphi_{E_0,L_1}(x,t_y)|$ (the thin blue lines) with the OBC eigenstate $|\psi_{E_0,G_{sq}}(x,y)|$ (the thick red lines) for different values of $t_y=y$. 
	We can find that they match exactly for small values of $t_y=y$;  
	while for large $t_y=y$, they only show deviations near the edges at $x=1$ and $x=N_x=80$. 
	This difference arises from the fact that we only consider the OBC wavefunction components that involve the $L_1$-edge, as shown in Fig.~\ref{F1}(d), where the transparency in blue represents the magnitude of the dynamical-duality wavefunction $\varphi_{E_0,L_1}(x,t_y)$.
	
	When we apply the above method to other edges, we can obtain $\varphi_{E_0,L_{2/3/4}}(x,t_y)$. 
	Finally, the constructed complete dynamical-duality wavefunction is given by:
	\begin{equation}
		\varphi_{E_0,G_{sq}}(x,t_y)=\sum_{i=1}^4 \varphi_{E_0,L_i}(x,t_y),
		\label{phi}
	\end{equation}
	which is plotted in the inset of Fig.~\ref{F1}(d), where the color transparency represents the dynamical-duality wavefunction $|\varphi_{E_0,G_{sq}}(x,t_y)|$. 
	The comparison between it and the numerical result $|\psi_{E_0,G_{sq}}(x,y)|$ in Fig.~\ref{F1}(b) shows the prefect match. 
	This demonstrates that we can indeed use the dynamical-duality wavefunction $\varphi_{E_0,G_{sq}}(x,t_y)$ to construct and understand the OBC eigenstate $\psi_{E_0,G_{sq}}(x,y)$ without losing any information.
	
	{\em {\color{red}GBZ of $E_0$.}}---We now examine the GBZ of $E_0$ under the square geometry, denoted as $\beta_{E_0,G_{sq}}$. 
	The validity of Eq.~\ref{phi} implies that
	\begin{equation}
		\beta_{E_0,G_{sq}}=\beta_{E_0,L_1}\cup\beta_{E_0,L_2}\cup\beta_{E_0,L_3}\cup\beta_{E_0,L_4},
	\end{equation}
	where $\beta_{E_0,L_i}$ represents the sub-GBZ for the $L_{i}$-edge. 
	We will now calculate $\beta_{E_0,L_1}$ as an illustrative example. 
	As detailed in Appendix, the dynamical-duality wavefunction for the $L_1$-edge constructed in this work has the following form
	\begin{equation}
		\varphi_{E_0,L_1}(x,y)=\sum_{k_x}\sum_{\alpha_{in}}A_{E_0,L_1}^{\alpha_{in}}(k_x)\left(e^{ik_x}\right)^x\left[z_{y,E_0}^{\alpha_{in}}(k_x)\right]^{y}, 
		\label{phinew}
	\end{equation}
	where $k_x=2\pi m/N_x$ and the first summation runs from $m=1$ to $N_x$. For a given $k_x$, the second summation counts all the roots of the ChE, i.e. $E_0-H_0(e^{ik_x},z_y)+i0^+=0$, enclosed by the unit circle $|z_y|=1$ in the complex-$z_y$ plane. 
	
	Firstly, in Eq.~\ref{phinew}, $\left(e^{ik_x}\right)^x$ is the only term that contains the variable $x$. 
	Comparing Eq.~\ref{phinew} with the following ansatz solution
	\begin{equation}
		\psi_{E_0}(x,y)=\sum_{(z_x,z_y)\in F_H(E_0)}A_{E_0}(z_x,z_y)\left(z_x\right)^x\left(z_y\right)^y, 
		\label{WFOBC1}
	\end{equation}
	we can find that $\mu_x$ must be zero for the $L_1$-edge. 
	Solving the ChE with $\mu_x=0$, we obtain two roots in our model, which is ordered by their absolute values, i.e. 
	\begin{equation}
		|z_{y,E_0}^{\alpha_1}(k_x)|\leq |z_{y,E_0}^{\alpha_2}(k_x)|
	\end{equation}
	The blue/red lines in Fig.~\ref{F2} (a) and (b) show the $\mu_{y,E_0}^{\alpha_1}(k)/\mu_{y,E_0}^{\alpha_2}(k)$ and $k_{y,E_0}^{\alpha_1}/k_{y,E_0}^{\alpha_2}(k)$, respectively. 
	
	Secondly, in Eq.~\ref{phinew}, $\left[z_{y,E_0}^{\alpha_{in}}(k_x)\right]^{y}$ is the only term that contains the variable $y$. 
	From the constraint of the second summation in Eq.~\ref{phinew}, i.e. the roots enclosed by the unit circle $|z_y|=1$ for given $k_x$, we require that 
	\begin{equation}
		\mu_{y,E_0}^{\alpha_{1}}(k_x)\leq0~{\rm and}~\mu_{y,E_0}^{\alpha_{2}}(k_x)\leq0,
		\label{mu}
	\end{equation}
	which is shown in Fig.~\ref{F2}(a) and (b) with opaque arcs. 
	
	Finally, in Eq.~\ref{phinew}, the term $A_{E_0,L_1}^{\alpha_{in}}(k_x)$ represents that superposition coefficient, whose absolute value is represented by the color transparency of the points/arrows in Fig.~\ref{F2}(b).  
	From Fig.~\ref{F2}(b), one can find that all the vectors are pointing to the $L_1$-edge, indicating that the wavefunction must be localized at the corresponding edge. 
	
	{\em {\color{red}Asymptotic GBZ of $E_0$.---}}~Note that $\mu_{y,E_0}^{\alpha_{1/2}}(k_x)$ represents the localization length near the $L_1$-edge.
	It is expected that the asymptotic behaviors of the wavefunction in the bulk are dominated by the root $z_{y,E_0}^{\alpha_2}(k_x)$, i.e., the opaque red curves in Fig.~\ref{F2} (a) and (b).
	When this approximation is taken, the corresponding GBZ is referred to as the asymptotic GBZ, denoted by $\beta^a$. 
	Fig.~\ref{F2}~(c) shows the asymptotic sub-GBZ of the $L_1$-edge, and Fig.~\ref{F2}~(d) shows the asymptotic GBZ of the square geometry. 
	The asymptotic sub-GBZs for different edges are labeled by different colors.
	
	Notably, it is observed in Fig.~\ref{F2}(d) that $\beta_{E_0,L_i}^a$ is constituted by a set of analytic arcs terminated at the Fermi points. 
	These analytic arcs are termed non-Bloch equal frequency contours. 
	When the edge direction changes, these non-Bloch equal frequency contours and their associated vectors change accordingly. 
	However, their endpoints, i.e., the Fermi points of $E_0$, remain invariant. 
	This key observation explains the the following puzzle in our numerical calculations, i.e., the coverage region of the OBC spectrum does not depend on the OBC geometry and further coincides with the coverage region of the PBC spectrum.
	
	\emph{\clr Generalizations.---}~Besides the above model, it is widely observed~\cite{Kai2022NC} and recently discussed in Ref.~\cite{HPHu2023} that the coverage region of the OBC spectrum $\sigma^{\rm OBC}$ is independent of the OBC geometry for different models. 
	Therefore, it is reasonable to expect that the physical picture of our model should be universal. 
	In general, based on the spectrum, we have two different classes of NHSE:
	(i) The generalized-reciprocal skin effect (GRSE), which satisfies
	\begin{equation}
		{\rm GRSE:}~~\sigma^{\rm PBC}=\sigma^{\rm OBC}=\sigma^{\rm OBC}_{G_i}=…=\sigma^{\rm OBC}_{G_j};
	\end{equation}
	(ii) The non-reciprocal skin effect (NRSE), which satisfies
	\begin{equation}\label{NRSESpectra}
		{\rm NRSE:}~~\sigma^{\rm PBC}\neq\sigma^{\rm OBC}=\sigma^{\rm OBC}_{G_i}=…=\sigma^{\rm OBC}_{G_j}.
	\end{equation}
	Here, $G_i$ and $G_j$ represent different open-boundary geometry shapes and the equal sign indicates the same spectral coverage region. 
	
	For the GRSE, since the PBC spectrum covers the same region with the OBC spectrum, it is natural to expect that the corresponding asymptotic GBZ is constituted by a set of geometry-independent Fermi points, and the corresponding geometry-dependent non-Bloch equal frequency contours connecting them. 
	
\begin{figure}[t]
		\begin{center}
			\includegraphics[width=1\linewidth]{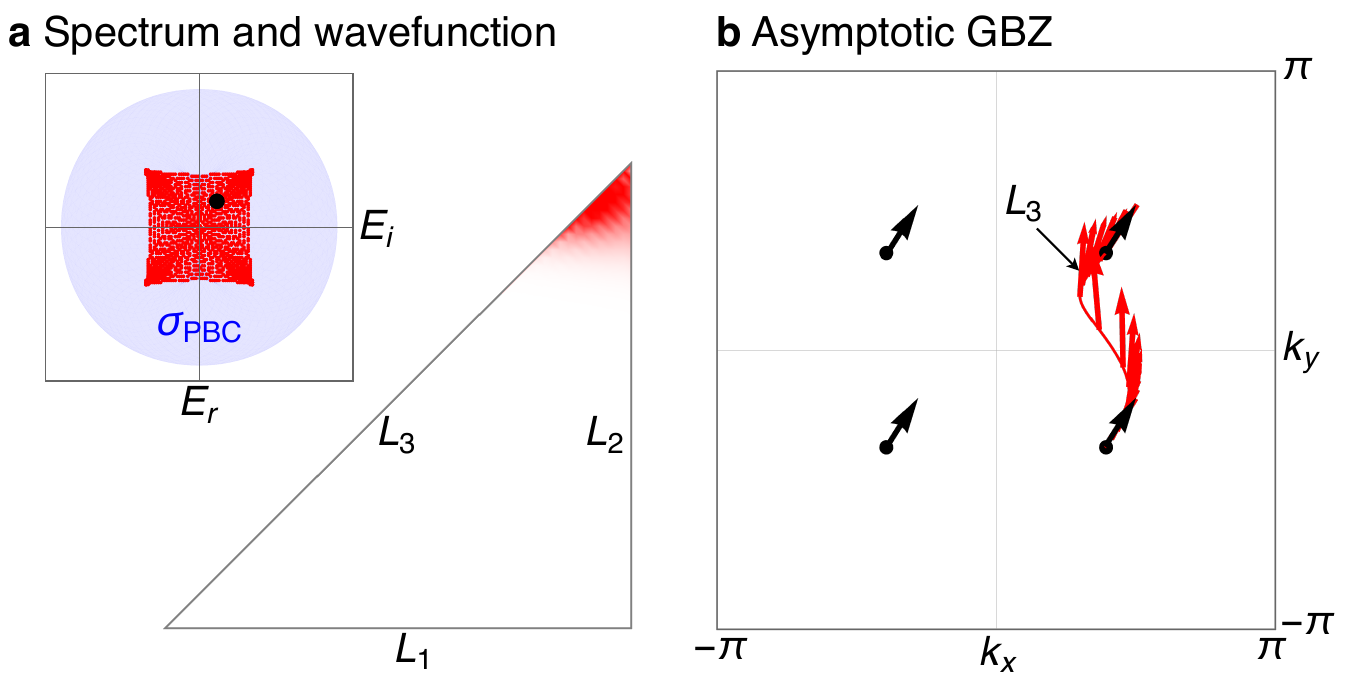}
			\par\end{center}
		\protect\caption{(a) depicts the energy spectrum and the wave function for a NRSE under triangle geometry. The PBC spectrum is represented by the light blue region, denoting $\sigma^{PBC}$, while the red dots correspond to $\sigma^{OBC}_{G_{tri}}$ in a triangle geometry. The chosen black dot indicates the energy eigenvalue $E_0=0.68+0.93i$, associated with the depicted wave function eigenstate.
			(b) displays the corresponding 2D asymptotic GBZ. Black points mark the non-Bloch Fermi points with nonzero arrows. The red lines trace the non-Bloch equal frequency contours of the $L_3$-edge, and the red arrows define the non-Bloch localization vector. 
			Here we note that there is no equal frequency contour for the $L_1$- and $L_2$-edges.} 
		\label{F3}
	\end{figure}

	For the case of NRSE, since the PBC spectrum is inconsistent with the OBC spectrum, the formula of asymptotic GBZ needs to be generalized. 
	Moreover, the invariant spectral coverage under different open-boundary geometries suggests the existence of a set of geometry-independent non-Bloch Fermi points. 
	These non-Bloch Fermi points can be connected by geometry-dependent non-Bloch equal frequency contours. 
	Now we use an example to demonstrate this point. 
	Consider the following Hamiltonian
	\begin{equation}\label{NRSEHam}
		H_2(z_x,z_y) = t_1 z_x + t_2/z_x + i t_3 z_y + i t_4/z_y 
	\end{equation}
	with $t_1=1/t_2=1/2, t_3=1/t_4=1/3$. 
	The PBC spectrum is shown in Fig.~\ref{F3} (a) with light blue color. 
	This model can be analytically solved under the square geometry using the separation of variables method. 
	Consequently, the OBC spectrum can be obtained as $E(k_x,k_y) = E_x(k_x)+E_y(k_y)=2\sqrt{t_1t_2}\cos k_x+i2\sqrt{t_3t_4}\cos k_y$, where $k_x,k_y\in \text{BZ}$. 
	For a given OBC eigenvalue $E_0$, the corresponding GBZ is constituted by four non-Bloch Fermi points, represented by four black vectors in Fig.~\ref{F3} (b). 
	
	However, with other types of open-boundary geometry, for example, the triangle geometry as shown in Fig.~\ref{F3} (a), $E(k_x,k_y)$ cannot be separated as $E_x(k_x)+E_y(k_y)$, and therefore the model can no longer be solved exactly by the separation of variable method. 
	As shown in Fig.~\ref{F3} (a), the red dots show the corresponding OBC spectrum under the triangle geometry, labeled by $\sigma_{G_{tri}}^{\rm OBC}$, which is inconsistent with the PBC spectrum, expressed as, $\sigma^{\rm PBC}\neq \sigma_{G_{tri}}^{\rm OBC}$; and a typical wavefunction for $E_0$ (the black dot) is localized at the corner of the triangle geometry. 
	To obtain the asymptotic GBZ for $E_0$ under the triangle geometry, we note that under the following imaginary gauge transformation:
	$z_x=\sqrt{t_2/t_1}\tilde{z}_x, \, z_y= \sqrt{t_4/t_3}\tilde{z}_y$, \
	the Hamiltonian in Eq.~\ref{NRSEHam} is transformed into
	\begin{equation}
		\tilde{H}_2(\tilde{z}_x,\tilde{z}_y) =  \sqrt{t_1 t_2} \, (\tilde{z}_x + \tilde{z}_x^{-1}) + i \sqrt{t_3 t_4} \, ( \tilde{z}_y + \tilde{z}_y^{-1}),
	\end{equation}
	which is a GRSE Hamiltonian and can be solved by the dynamical-duality method proposed in this work.  
	Fig.~\ref{F3} (b) shows the corresponding 2D asymptotic GBZ of $E_0$ under the triangle geometry. 
	We show that the asymptotic GBZ for $L_3$-edge is constituted by four geometry-independent non-Bloch Fermi points (the four black arrows in Fig.~\ref{F3}(b) and geometry-dependent non-Bloch equal frequency contours (the set of red arrows); in contrast, the asymptotic GBZ for $L_1$- and $L_2$-edges only comprise four non-Bloch Fermi points without non-Bloch equal frequency contours. 
	Remarkably, we demonstrate that for the asymptotic GBZ of $E_0$, the non-Bloch Fermi points are independent of open-boundary geometries, such as the square or triangle geometries, thus extending the asymptotic GBZ to the case of NRSE. 
	
	\emph{\clr Discussions and conclusions.---}~In this Letter, we propose a theory for the 2D NHSE, which can not only be applied to describe the exact localization behaviors of the skin modes, but also can be applied to understand the OBC spectrum under different geometries. 
	One corollary of our theory is that $\sigma^{\rm OBC}$ should be insensitive to the OBC geometry shape, a phenomenon widely observed in the literature. 
	Another corollary of our theory is that if the Hamiltonian possesses inversion or anomalous spinless time-reversal symmetry, the corresponding system in general belongs to the GRSE. 
	Consequently, the bulk-boundary correspondence should be preserved in most 2D non-Hermitian systems that exhibit these symmetries under arbitrary OBC geometry $G_0$. 
	Furthermore, the corresponding symmetry-protected exceptional semimetals should also be robust to variations in OBC geometry $G_0$. 
	Similarly, the corresponding exceptional points should uphold the Fermi doubling theorem~\cite{YangFermiDoubling2021}. 
	
	Although our work have made a significant step towards establishing a general theory for the 2D NHSE, there are still several important questions should be clarified in the following studies. 
	(i) The dynamical-duality method proposed in this work to calculate the 2D GBZ can only be applied to the GRSE and specific NRSE models.
	The generalization of our method to generic NRSE is an important step in the forthcoming works. 
	(ii) For generic NRSE models, we believe that the non-Bloch Fermi points is related to the points where Ronkin function is zero proposed in the amoeba theory in Ref.~\cite{HYWang2022}. 
	Further numerical verifications should be studied in the following works. 
	
	\appendix
	\section*{Appendix}
	
	{\color{red} \em The formal definition of 2D GBZ.}---We begin with a general non-Bloch Hamiltonian  $H(z_x,z_y)$, where $(z_{x},z_y)\in\mathbb{C}^2$ represents the non-Bloch momentum and  can be expressed as: 
	\begin{equation}
		z_x=e^{ik_x+\mu_x},~~~~z_y=e^{ik_y+\mu_y}.
		\label{z}
	\end{equation}
	Here $(k_{x},k_y)$ is the Bloch momentum, and $(\mu_x,\mu_y)$ is the non-Bloch localization vector. 
	The length and direction of this vector determine the localization strength and direction, respectively.
	For a given OBC geometry $G_0$, we denote the OBC spectrum as $\sigma_{G_0}^{\rm OBC}$. 
	Then, for any given $E_0\in\sigma_{G_0}^{\rm OBC}$, the corresponding OBC eigenstate can be written as
	\begin{equation}
		\psi_{E_0,G_0}(x,y)=\sum_{(z_x,z_y)\in F_H(E_0)}A_{E_0,G_0}(z_x,z_y)\left(z_x\right)^x\left(z_y\right)^y.
		\label{WFOBC}
	\end{equation}
	In this equation, $A_{E_0,G_0}(z_x,z_y)$ is the linear superposition coefficient, which depends on $E_0$, $z_{x/y}$ and the OBC geometry $G_0$. 
	$F_H(E_0)$ is the solution of the ChE, defined as: 
	\begin{equation}
		F_H(E_0):=\{(z_x,z_y)\in\mathbb{C}^2|\det[E_0-H(z_x,z_y)]=0\}.
		\label{FH}
	\end{equation} 
	It is important to note that not all solutions that belong to $F_H(E_0)$ can contribute to the OBC eigenstate. 
	For instance, in the Hermitian case, only the solutions $(z_x,z_y)$ satisfying $(\mu_x,\mu_y)=(0,0)$ can have a nonzero $A_{E_0,G_0}(z_x,z_y)$. 
	In non-Hermitian systems, it is natural to ask what conditions $(z_x,z_y)$ must satisfy for $A_{E_0,G_0}(z_x,z_y)$ to be nonzero. 
	This leads us to define the 2D GBZ of $E_0$ (under the OBC geometry $G_0$) as:
	\begin{equation}
		\beta_{E_0,G_0}:=\{(z_x,z_y)\in \mathbb{C}^2|\lim_{N\rightarrow\infty}A_{E_0,G_0}(z_x,z_y)\neq0\}.
		\label{GBZE0}
	\end{equation}
	Our task is to calculate $\beta_{E_0,G_0}$ for any given $E_0$, $G_0$, and $H(z_x,z_y)$.  
	The specific values of $A_{E_0,G_0}(z_x,z_y)$ are beyond the scope of this work. 
	It is worth noting that the GBZ defined here is $E_0$-dependent, which differs from the 1D case where the GBZ is defined for the entire OBC spectrum.

	{\color{red} \em The calculation procedure of the dynamical-duality method.}---
	which is equal to the OBC eigenstate of $E_0$ at the $L_1$-edge, i.e.,
	\begin{equation}
		\varphi_{E_0,L_1}(x,t_{y}'):=\psi_{E_0,G_{sq}}(x,y=1),~{\rm where}~~t_{y}'=1.
		\label{WFini}
	\end{equation}
	
	Now we outline the calculation procedure for $\varphi_{E_0,L_1}(x,t_y)$:
	
	(i). First, take the Fourier transform for $\varphi_{E_0,L_1}(x,t_{y}')$ as shown in Eq.~\ref{WFini}, i.e.,
	\begin{equation}
		\phi_{E_0,L_1}(k_x,t_{y}')=\frac{1}{\sqrt{N_x}}\sum_{x=1}^{N_x}e^{-ik_xx}\varphi_{E_0,L_1}(x,t_{y}'),
		\label{ck}
	\end{equation}
	where $k_x=2\pi m/N_x$ with $m=1,...,N_x$. 
	
	(ii). For any given $k_x\in[-\pi,\pi]$ and $t_y>1$, use the residue theorem to calculate the propagator:
	\begin{equation}
		\begin{aligned}
			\relax g_{E_0,L_1}(k_x,t_y)
			&=\oint_{C_1}\frac{dz_y}{(iz_y)}\frac{(z_y)^{t_y}}{E_0-H_0(e^{ik_x},z_y)+i0^+}\\
			&=\sum_{\alpha_{in}}C^{\alpha_{in}}_{E_0}(k_x)\left[z_{y,E_0}^{\alpha_{in}}(k_x)\right]^{t_y},\\
		\end{aligned}
		\label{gk}
	\end{equation}
	where $C_1$ indicates the unit circle in the complex $z_y$ plane, $z_{y,E_0}^{\alpha_{in}}$ is the $\alpha_{in}$-th root of $E_0-H_0(e^{ik_x},z_y)+i0^+=0$ enclosed by $C_1$, and the summation counts all the roots that are inside the unit circle. 
	Note that this propagator is a geometry-independent quantity. 
	
	(iii). Finally, calculate $\varphi_{E_0,L_1}(x,t_y)$, i.e.,  
	\begin{equation}\begin{aligned}
			\varphi_{E_0,L_1}(x,t_y)=\frac{1}{\sqrt{N_x}}\sum_{k_x}e^{ik_xx}B_{E_0,L_1}(k_x) \, g_{E_0,L_1}(k_x,t_y),
			\label{phiL}
	\end{aligned}\end{equation}
	where $k_x=2\pi m/N_x$ and the summation counts $m$ from $1$ to $N_x$. 
	When $t_y=t_y'$, $\varphi_{E_0,L_1}(x,t_y)$ should reduce to $\varphi_{E_0,L_1}(x,t_y')$ appeared in Eq.~\ref{ck}, which requires that 
	\begin{equation}
		B_{E_0,L_1}(k_x)=\frac{\phi_{E_0,L_1}(k_x,t_{y}')}{g_{E_0,L_1}(k_x,t_{y}')}.
		\label{Coeff}
	\end{equation}
	It should be noted that our method can be generalized to other OBC with arbitrary polygon geometries. 
	
	By substituting Eq.~\ref{gk} into Eq.~\ref{phiL} and replacing $t_y$ by $y$, we obtain 
	\begin{equation}
		\varphi_{E_0,L_1}(x,y)=\sum_{k_x}\sum_{\alpha_{in}}A_{E_0,L_1}^{\alpha_{in}}(k_x)\left(e^{ik_x}\right)^x\left[z_{y,E_0}^{\alpha_{in}}(k_x)\right]^{y}, 
		\label{phinew1}
	\end{equation}
	where $k_x=2\pi m/N_x$, and the first summation runs from $m=1$ to $N_x$, and the second summation counts all the roots enclosed by $C_1$ for a given $k_x$. 
	The linear superposition coefficient, 
	\begin{equation}
		A_{E_0,L_1}^{\alpha_{in}}(k_x)=\frac{1}{\sqrt{N_x}}B_{E_0,L_1}(k_x)C_{E_0}^{\alpha_{in}}(k_x)
		\label{A}
	\end{equation}
	is a function of $k_x$ and $\alpha_{in}$ for a given $E_0$ and the selected $L_1$-edge.

	\bibliography{refs}

\begin{thebibliography}{73}
\bibitem [{SM1()}]{SM1}%
  \BibitemOpen
  \href@noop {} {\bibinfo  {journal} {Here we note that in this work, in order
  to have a good representation for the vector field, all the vectors are
  rescaled to its 1/2 length}\ }\BibitemShut {NoStop}%
\end{thebibliography}
	\bibliographystyle{apsrev4-1}
	
\end{document}